# Ultra-bright and energy-efficient quantum-dot LEDs by idealizing charge injection

Yizhen Zheng,[1]† Xing Lin,[2]†* Jiongzhao Li,[1] Jianan Chen,[1] Zixuan Song,[2] Yuan Gao,[3] Huifeng Wang,[1] Zikang Ye,[1] Haiyan Qin,[1] & Xiaogang Peng[1]*

[1]Key Laboratory of Excited-State Materials of Zhejiang Province, Department of Chemistry, Zhejiang University, Hangzhou 310027, China.

[2]Key Laboratory of Excited-State Materials of Zhejiang Province, College of Information Science and Electronic Engineering, Zhejiang University, Hangzhou 310027, China.

[3]Najing Technology Corporation Ltd., Hangzhou 310027, China.

†These authors contributed equally to this work.

*Email: lxing@zju.edu.cn; xpeng@zju.edu.cn

**Abstract:**

Lighting and display, relying on electric and optical down-conversion emission with sluggish power efficiency, account for >15% global electricity consumption[1,2]. In 2014, quantum-dot (QD) LEDs (QLEDs) with near-optimal external quantum efficiency emerged[3] and promised a pathway to avoid the vast down-conversion energy loss[4,5]. Despite a decade of progress[4-22], fabrication of energy-efficient QLEDs with application-relevant brightness remains elusive. Here, the main roadblock is identified as the oxidative species adsorbed in the nanocrystalline electron-injection layer of QLEDs, which is then addressed by a simple reductive treatment to simultaneously boosts electron conductivity and hole blockage of the electron-injection layer. The resulting sub-bandgap-driven QLEDs with optimal efficiency achieve ultra-high brightness across the entire visible spectrum at least 2.6-fold higher than existing benchmarks. The brightness fully satisfies the demands of various forms of lighting and display, which surges to a remarkable level sufficient for QD laser diodes with a moderate bias (~9 V). Optimized electron injection further enables new types of QD-blend LEDs for diffuse white-light sources surpassing the 2035 R&D targets set by the U.S. Department of Energy. Our findings open a door for understanding and optimizing carrier transport in nanocrystalline semiconductors shared by various types of solution-processed optoelectronic devices.

## Main Text:

### Ultra-bright and energy-efficient QLEDs

The energy efficiency of an LED is best captured by its external power efficiency (EPE) which is a composite of three components: the light-extraction efficiency (LEE), the internal quantum efficiency (IQE), and the ratio of the bandgap voltage ($V_p = h\nu/e$, with $h\nu$ as the average emitting photon energy and $e$ as elementary charge) to the driving voltage $V$. Alternatively, EPE can be expressed as the product of either EQE and $V_p/V$ ratio or LEE and IPE:

$$\text{EPE} = (\text{LEE} \times \text{IQE}) \times V_p/V = \text{EQE} \times V_p/V = \text{LEE} \times \text{IPE}$$

Though the LEE of planar LEDs is typically 20~30% (Extended Data Fig. 1), it can reach ~90% by additional optical engineering[23,24]. Lighting and display currently rely on GaN quantum well LEDs coupled with phosphors and/or organic LEDs. Albeit with optimized LEE, both technologies share the same drawback at the luminance levels needed for lighting and display, namely, the electrical and optical down-conversion emission with their $V_p/V$ ratios much below one, implying an intrinsically low EPE.

QLEDs in a common structure (Fig. 1a), with nanocrystalline zinc-magnesium-oxide (ZnMgO) as the electron-injection layer (EIL)[12,25,26], are fabricated with a quasi-monolayer of close-packed CdSe-based core/shell QDs as the emitting layer (Fig. 1a and Extended Data Fig. 1)[5,19,27]. In literature, the performance of QLEDs is reported to be universally determined by the encapsulation resins, i.e., resins containing volatile acids (acid-containing resin) versus common acid-free resins[28]. High EQE QLEDs in the literatures are encapsulated with the acid-containing resin, but their EQE and current density need to be improved by shelf storage—known as positive aging (Extended Data Fig. 2-3)[28-32]—likely via unpredictable and uncontrollable reaction(s) between the volatile acids and the nanocrystalline EIL. Positive aging may also occur for the QLEDs operating under a constant-current mode[33], followed by destructive effects possibly caused by reaction(s) between the volatile acids and the metal electrode[28]. The unpredictable and lengthy positive aging in QLEDs across the visible spectrum, poses an unbearable challenge for commercialization of QLEDs. Alternatively, QLEDs encapsulated with the acid-free resin typically exhibit low EQE and minimum current density (Extended Data Fig. 2-3). Consistent with their problematic optoelectronic properties, Fig. 1b (inset *C1* and *C2*) shows that QLEDs encapsulated with either type of resin present poor ideality-factors ($n_J$) in terms of charge injection[34].

Our new fabrication scheme excludes the acid-containing resin and features a water-vapor treatment after deposition of the top metal electrodes. Typically, following deposition of the aluminum electrode (~70 nm), the device—prior to encapsulation—is transferred into an atmosphere-controlled chamber for exposure to a water-vapor/argon atmosphere (1 bar, 50% relative humidity, ~30 minutes, see Methods). The current density of the resulting QLEDs reaches a record-high level and their ideality-factors under bias below $V_p$ are greatly improved (Fig. 1b, inset *T*), indicating near-optimal charge injection.

Operating at $V_p$ (1.93 V) (Fig. 1c, inset), a typical red-emitting (644 nm) QLED achieves a luminance as high as 2,900 cd/m² and a peak EQE of 28.9% (Fig. 1b-c). A superior peak EPE (29.1%) is attained at 2,100 cd/m$^2$ (Fig. 1c). At a driving voltage of 3 V, the luminance surges to 94,000 cd/m$^2$. With an LEE of 32% (Extended Data Fig. 1), the peak IPE surpasses 90%, and the IPE across a luminance range of 500-10,000 cd/m$^2$ is sustained above 81%. Our devices exhibit outstanding operational- and shelf-stabilities, contrasting with positive-aging typically observed in conventional QLEDs[28,33]. Post shelf-storage for varied periods, the QLEDs maintain nearly constant EQE, luminance, current density, emission uniformity (Fig. 1d), and ideality-factor (Extended Data Fig. 4). With a constant current density of 150 mA/cm$^2$ and an initial luminance of approximately 30,000 cd/m$^2$, the red-emitting (644 nm) device demonstrates a long $T_{95}$ (111 hours) and $T_{80}$ lifetime (450 hours), revealing a minimal initial luminance increase and a sub-linear rise in voltage over time (Fig. 1e). Results in Fig. 1f demonstrate high reproducibility of the new scheme, statistically 90.6 ± 3.1% IPE at 1,000 cd/m$^2$ for a batch of 31 red-emitting (644 nm) QLEDs.

**Energy-efficient QLEDs for all colors**

The simple water-vapor treatment is readily extended to QLEDs with other emitting colors, resulting in highly efficient and bandgap-driven QLEDs across the entire visible spectrum with their luminance levels 2.6 to 12.3 times higher than the benchmark records (Fig. 2a-c) and free of positive-aging effects (Extended Data Fig. 2-3). Driven at the bandgap (or sub-bandgap), performance of the water-treated devices across the visible spectrum are drastically advanced in comparison with their counterparts encapsulated with the acid-free resin without water-treatment, specifically, 1 to 2 orders of magnitude increase of luminance, 1.5 to 8 times of EQE, 1 to 3 orders of magnitude increase of lifetime (Extended Data Fig. 2 and 3). Importantly, the shorter the emitting wavelength is, the more dramatic the effects of the water treatment would be.

Consistent with their near-ideal electronic properties across the visible spectrum (Extended Data Fig. 4), luminance of efficient QLEDs driven to ~1.5 times of $V_p$ increases exponentially to a level sufficient for various lighting and display applications without considering additional light extraction (Fig. 2d-f). For instance, the water-treated green- and red-emitting QLEDs can readily match the brightness requirements of the augmented and virtual reality displays yet universally offer >20% EPE, superior to the quantum-well micro-LEDs with single-digit EPE[35].

Further increasing the bias to ~10 V, especially with pulsed excitation to partially alleviate thermal damage under high current density[36], transient peak luminance levels of the QLEDs on glass substrates reach record-high maxima (red 12,400,000 cd/m$^2$, green 23,800,000 cd/m$^2$, and blue 1,180,000 cd/m$^2$ in Fig. 2d-f). These transient luminance levels approach the population-inversion limit for realizing QD laser diodes[18]. For example, when the luminance of the red-emitting QLEDs exceeds ~5,000,000 cd/m$^2$ (biased at 9 V), a high-energy peak (85 meV above the band-edge electroluminescence) emerges (Extended Data Fig. 5), which closely resembles that excited by a pulsed laser at power density sufficient to observe amplified spontaneous emission (Extended Data Fig. 5). By minimizing the loss associated with the electrodes and enhancing the modal gain, it is anticipated that electrically pumped QD laser diodes will be achieved.

By integrating a hemispherical lens to reduce the optical waveguide losses only, the certified EPE values (see Supplementary Information) for red-emitting, green-emitting, and blue-emitting QLEDs stand at 55%, 46%, and 32%, respectively. This enhancement in LEE (1.6-1.9 times) achieved via hemispherical lenses aligns well with both our calculations and literature reports[23].

**Nature of the oxidative trap states**

A water-vapor treatment results in a negligible impact on every layer deposited prior to the ZnMgO deposition (Extended Data Fig. 6), confirming the electron traps within the ZnMgO layer. Oxidative species such as oxygen molecules are known as common electron traps for the ZnMgO EIL[31,37-39]. As an *n*-type semiconductor, ZnMgO's electrical conductivity is known to arise from O-vacancy sites that should be inherently sensitive to ambient oxygen[40,41]. These facts urge us to carry out *in-situ* studies of the water-vapor and oxygen treatments on the unencapsulated QLEDs (see Methods).

During the continuous water-vapor exposure, the current density at 3 V ($J_{3V}$) of a green-emitting (532 nm) QLED rises gradually from 8 to 105 mA/cm$^2$ (Fig. 3a, upper panel). The luminance at 3 V ($L_{3V}$) follows a similar trend but with an accelerated pace, leading to an early reached EQE plateau (Fig. 3a, bottom). During this initial water-vapor treatment (Fig. 3b, top), the current density increases sharply above ~0.6 V yet reduces below this voltage, signifying improved electron conductivity and hole blockage of the ZnMgO layer.

Conversely, a water-vapor treatment before the aluminum deposition increases both the current density and luminance but barely improves the EQE. This deficiency can be improved by additional water-vapor treatment after the aluminum deposition, which triples the EQE (> 23%), doubles the luminance at 3 V (Extended Data Fig. 6), and suppress the current density below ~2 V yet retaining the high current density above ~2 V (Fig. 3b, top inset). These striking contrasts between two types of water treatments are qualitatively confirmed for the HTL-emitter LEDs (Extended Data Fig. 6).

The results in the above two paragraphs hint at the presence of two distinct types of traps. The weakly adsorbed ones—shallow traps—only reduce electron conductivity of the ZnMgO layer and could be displaced by the water vapor alone. The strongly adsorbed ones are deep traps causing hole leakage of the ZnMgO layer[42] and could only be removed by the highly reductive species, such as hydrogen radicals[43] generated by the reaction between water and aluminum. Since both types of traps can be efficiently removed by the *in-situ* generated reductive species, they should be oxidative species, which are likely oxygen-related species accumulated during the synthesis and processing of the ZnMgO nanocrystals. This hypothesis is supported by *in-situ* studies on an oxygen (1% oxygen/nitrogen mixture) treatment of the fully water-treated QLEDs.

Figure 3a-b illustrate that an oxygen treatment rapidly results in almost regeneration of the poor performance of the pristine QLED, namely, extremely low $J_{3V}$, merely observable $L_{3V}$, and a minimum EQE. Similar to the original electron traps in the pristine QLEDs, all electron traps

created by the oxygen treatment can be completely removed by subsequent water-vapor treatment with the top electrode. Figure 3a further reveals that the oxygen-water treatment cycle is fully reproducible. However, an extended oxygen treatment would create abundant weakly and strongly adsorbed electron traps (Extended Data Fig. 7), leading to a slower water-induced recovery closely resembling that in the initial water treatment for the pristine QLED.

The standard water-vapor treatment is consistently effective for the EIL composed of either zinc-oxide or zinc-magnesium-oxide nanocrystals synthesized via different routes (see Methods) (Extended Data Fig. 8). The water-vapor treatment on the QLEDs with a titanium top-electrode is similarly effective with those with an aluminum electrode (Extended Data Fig. 8). For the QLEDs with a silver electrode, the water-vapor treatment affects the device in a manner similar to the QLEDs without any top electrode, presumably because of the poor reductivity of the silver metal. However, if the devices are biased beyond ~2.6 V repeatedly for ~6 $J$-$V$ sweeps (Extended Data Fig. 8), both electron conductivity and hole blockage of the same devices would resemble those of typical QLEDs. Similar to the electrochemical reduction of zinc-carboxylate (or cadmium-carboxylate) ligands in QLEDs[44], these voltage-induced improvements are presumably due to electrochemical reduction of water to generate active hydrogen radicals on the silver nanostructures[45].

Electrochemical reduction also occurs during the oxygen treatment of QLEDs. When the oxygen treatment (see above) is carried out without any electrical bias yet with a high oxygen pressure (1 bar), performance of the QLEDs degrades very slowly (Extended Data Fig. 9). Furthermore, the adverse effects of the power-off oxygen treatment can be largely eliminated by simply storing the devices for several days (Extended Data Fig. 9). These findings suggest that, akin to the original electron traps, the oxygen-related species also include two types. The weakly adsorbed ones are likely physically adsorbed oxygen molecules on the ZnMgO nanocrystals, exacerbating its electron conductivity, and can be removed either through competitive adsorption by water or through spontaneous desorption. The electrochemical reduction of oxygen molecules during device operation should generate the negatively charged oxygen-related species (e.g. $O_2^{-\cdot}$), which could strongly be chemically adsorbed onto the positively charged surface metal sites on the nanocrystalline oxides and are only removable through the reactions with the reductive species generated either chemically or electrochemically. These deep traps act as electron-hole recombination centers, assisting the hole leakage and deteriorating the EQE (Extended Data Fig.6).

The electroluminescence spectra of a QLED after different treatments align with each other completely, though the oxygen-treatment induces reversible dark spots in the electroluminescence image (Fig. 3c). This indicates that there are no structural or optical damages to the QDs by either native or oxygen-induced electron traps. Conversely, the visible trap photoluminescence of the ZnMgO layer[38,39] is activated by the oxygen treatment, while its ultraviolet band-edge photoluminescence[31,37] is suppressed (Fig. 3d). In contrast, the water-vapor treatment after the aluminum deposition induces an opposite change for both the pristine and the oxygen-treated ZnMgO layer. Being less effective, a water-vapor treatment without a top electrode can also partially increase (suppress) the ZnMgO band-edge (trap) photoluminescence (Fig. 3d, inset), consistent with existence of two types of oxidative species on the ZnMgO layer of either pristine or oxygen-treated QLEDs. Again, the similarities between the pristine ZnMgO layer and the oxygen-treated one revealed in Fig. 3d strongly suggest that the electron traps in the former must be oxidative species and most probably oxygen-related.

**Broadband QLEDs**

Solid-state lighting would greatly benefit from "broadband QLEDs" ($\beta$QLEDs), fabricated with a monolayer of QD blend with various emitting photon energies[46,47]. Fig. 4a demonstrates that, at a practical current density (1 mA/cm$^2$), the driving voltage of our QLEDs is linearly related to yet ~0.15 V below its $V_p$, indicating near-ideal electron injection of sub-bandgap-driven QLEDs across the visible spectrum.

A prototypical $\beta$QLED, utilizing a blend of QDs emitting at 605, 623, and 644 nm and operated at 1.90 V—below the average photon voltage—yields electroluminescence with a 55 nm full-width-half-maximum, a luminance of 1050 cd/m$^2$, and an impressive IPE of 91.9% (Fig. 4b). Evaluated by Commission Internationale de l'Eclairage (CIE) *x* and *y* coordinates, minimal color shift across a luminance range of 100-10,000 cd/m$^2$ (Fig. 4c) and negligible color drift between 0 and 100 hours of continuous operation at a constant voltage (Extended data Fig. 10) are observed, consistent with the uniform electron-injection efficiencies across the visible spectrum (Fig. 4a).

By employing three βQLEDs operating in the orang-red, green-yellow, and blue-green spectral windows, it becomes possible to create a continuous-spectrum white-light source closely emulating the black-body radiation at 3500 K, achieving an excellent color rendering index (CRI) of 90.4 (Fig. 4d). With the certified EPEs for the monochromatic QLEDs coupled with semispherical lenses, luminous efficacy of this warm-white QLED reaches 168-181 lm/W within the intensity range of 5,500-55,000 lm/m$^2$ (with theoretical upper limit estimated to be 353 lm/W, see Methods), which corresponds to ~50% EPE. This efficacy and CRI surpass the R&D goals for diffuse light sources set by the U.S. Department of Energy for the year 2035 (CRI ≥ 80 and efficacy ≥ 150 lm/W)[48].

## Summary and outlook

A simple water-vapor treatment of QLEDs with the top metal electrode converts the ZnMgO EIL to excel in hole blockage yet markedly facilitate electron injection/transport than typical organic EILs[49,50]. The outstanding charge-injection/transport properties of the optimized nanocrystalline EIL promote the QLEDs driven near their bandgaps to be unique electroluminescence devices with near-unity internal power efficiency and ultra-high brightness. Results here further reveal that, the higher the emitting photon energy is, the much greater the improvements by the optimized nanocrystalline EIL are, implying a new direction for eventually unifying the performance of the blue-, green-, and red-emitting QLEDs. The exceptional performance combined with the low-cost and scalable solution processing of QLEDs open up new horizons for developing advanced lighting, display, and lasing technologies.

# Main Text References

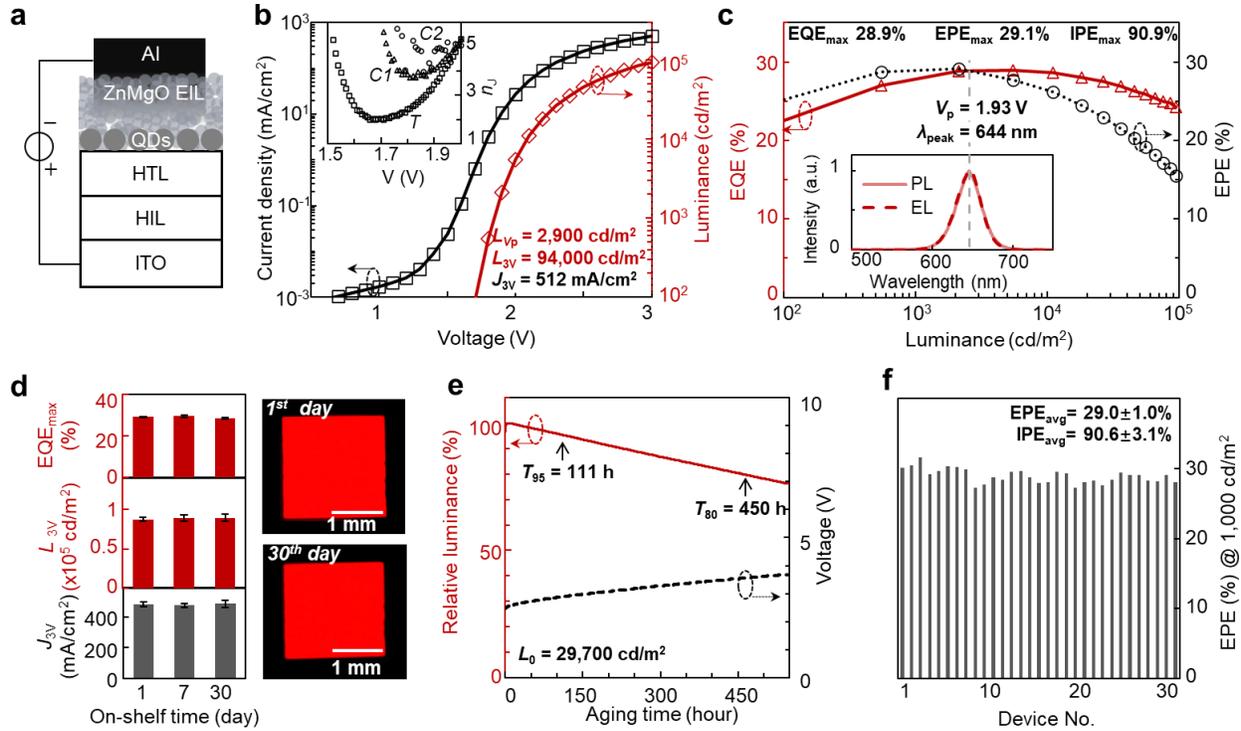

**Fig. 1| Ultra-bright QLEDs with near-unity internal power efficiency (IPE). a**, Schematic of layers in the device structure (indium-tin-oxide (ITO)/hole-injection layer (HIL)/hole-transport layer (HTL)/QD/electron-injection layer (EIL)/aluminum). **b**, Current density-voltage-luminance (*J-V-L*) characteristics of a red-emitting (644 nm) QLED. Inset: current ideality-factor ($n_J$) versus voltage (*V*) for control device *C1* encapsulated with the acid-free resin, *C2* encapsulated with the acid-containing resin, and the typical device *T* with the water-vapor treatment encapsulated with the acid-free resin. **c**, EQE and EPE versus luminance for the device shown in **b** in a broad luminance range. Inset: photoluminescence (PL, solid line) of the QD solution and electroluminescence (EL, dashed line) of a QLED. The crossing point of EPE and EQE curves corresponds to the average energy of emitted photons (1.93 eV). **d**, Left: peak EQE (EQE$_{max}$), luminance ($L_{3V}$) and current density ($J_{3V}$) at 3 V versus shelf-aging time. Error bars represent the standard deviations derived from one batch of four typical devices. Right: the photographs of a working QLED on the 1$^{st}$ and 30$^{th}$ day. **e**, Relative luminance and voltage versus operational time (driven at 150 mA/cm$^2$). **f**, EPE at 1,000 cd/m$^2$ from 31 devices.

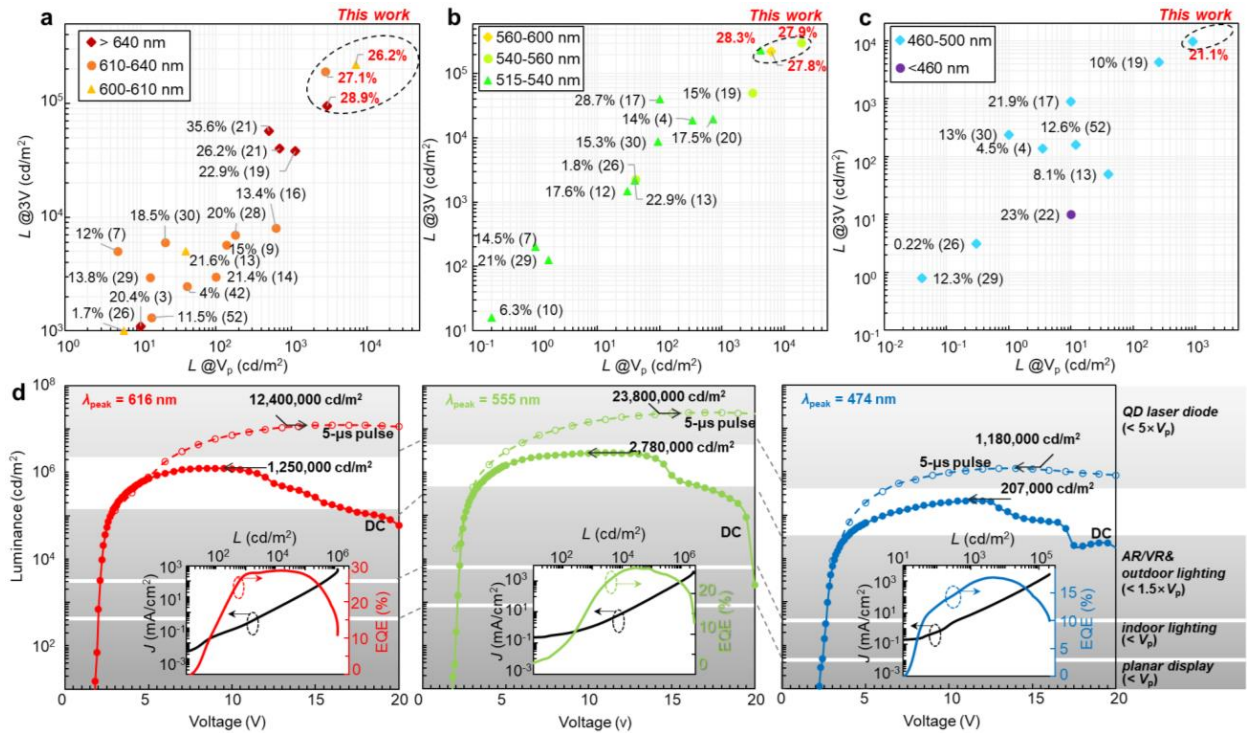

**Fig. 2| Ultra-bright and energy-efficient QLEDs across the visible spectrum. a-c**, Comparative analysis of our top-performing QLEDs against previously reported high-performing QLEDs in the orange-red (**a**), green-yellow (**b**), and blue-cyan (**c**) spectral windows, respectively. The *x*- and *y*-axis in each plot is the luminance at the corresponding $V_p$ and at a 3 V bias. The peak EQE values are annotated next to each data point, with the respective reference numbers in parentheses. **d**, Voltage-luminance characteristics of red-, green- and blue-emitting QLEDs under DC or pulsed excitation (solid line: 4 mm$^2$, dashed line: 0.02 mm$^2$ emitting area). The shaded areas indicate the luminance levels required for different scenarios outlined to the right. Inset, current density-luminance-EQE characteristics under DC bias.

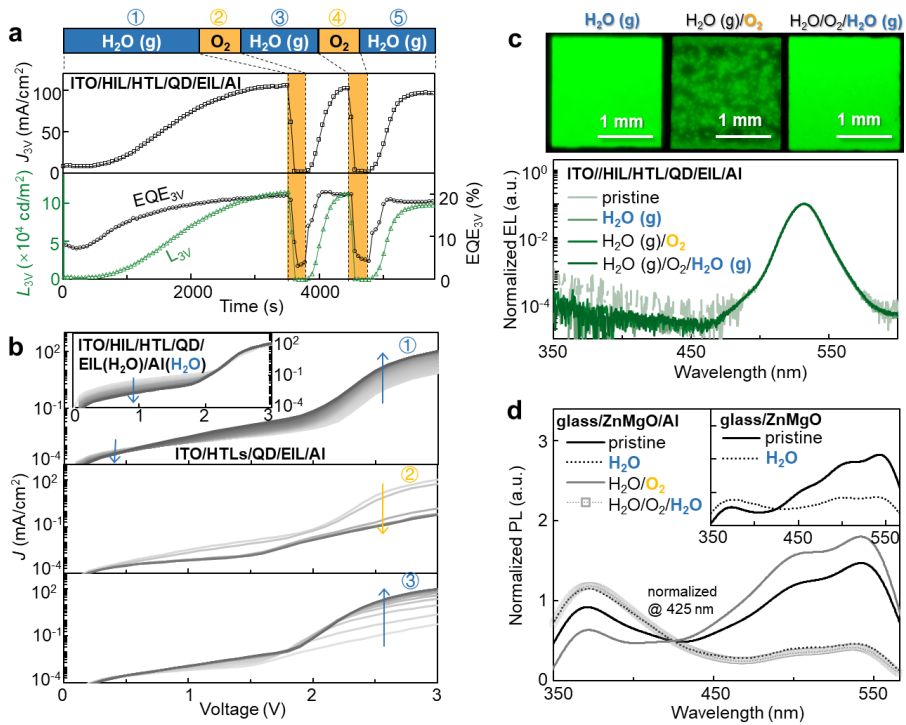

**Fig. 3| Formation and elimination of electron traps in a ZnMgO electron-injection layer. a**, Temporal evolutions of *in-situ* monitored luminance ($L_{3V}$), current density ($J_{3V}$), and EQE at 3 V of a green-emitting QLED exposed to the atmospheres indicated by the text-box above. $H_2O$ (g) stands for water-vapor. **b**, *In-situ* current density (*J*) versus voltage (*V*) for the typical device treated with the initial water-vapor treatment with the top aluminum electrode (top panel), a low-concentration of oxygen (middle panel), and the second water-vapor treatment after the oxygen treatment (bottom panel). Inset: temporal evolution of *J-V* characteristics during the water-vapor treatment with the top aluminum electrode of a device treated with water-vapor prior to deposition of the top aluminum electrode. **c**, Photographs (top panel, driven at 3.0 V) and normalized EL spectra of QLEDs (bottom panel, driven at 4.0 V) treated with different atmospheres. **d**, The normalized PL spectra of the glass/ZnMgO/Al sample treated with different atmospheres. Inset: the normalized PL spectra of the glass/ZnMgO sample with and without the water-vapor treatment.

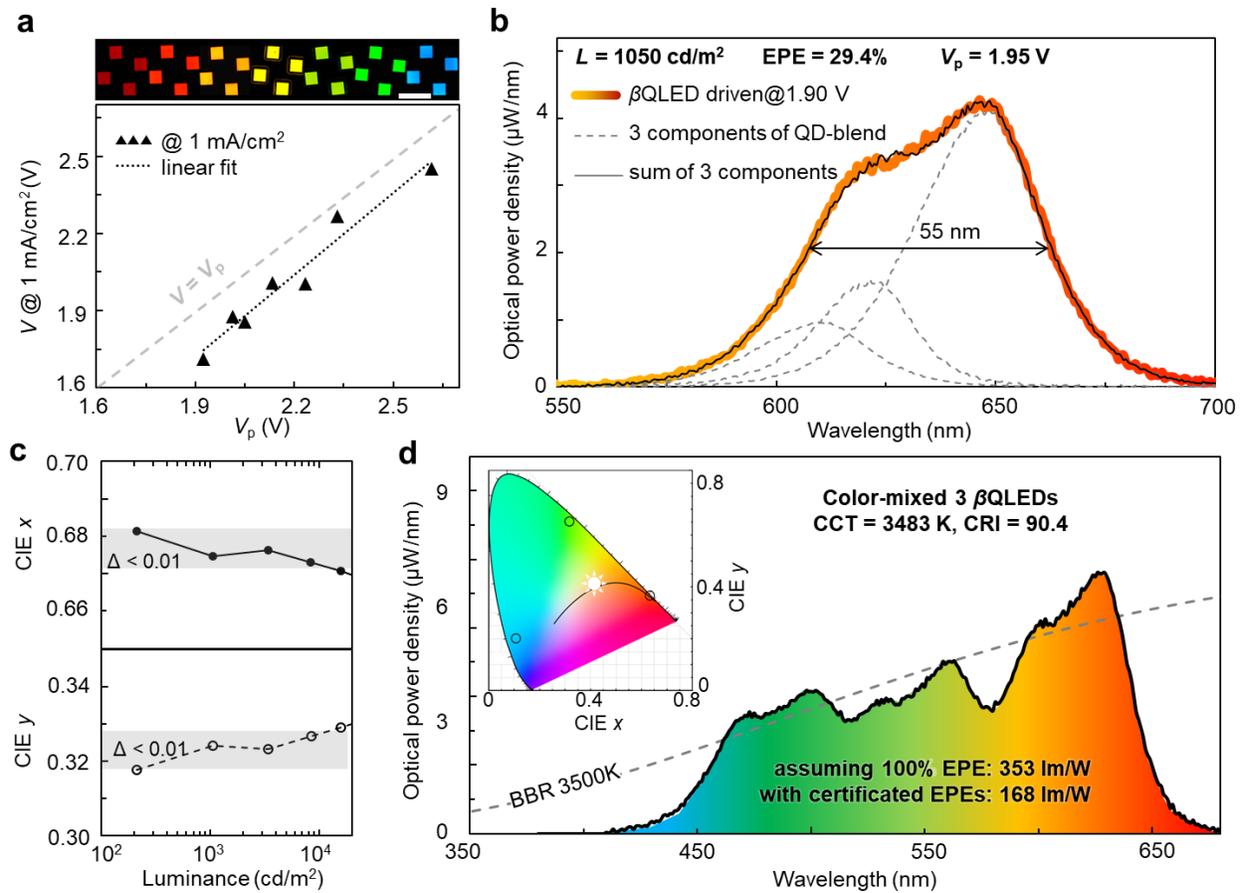

**Fig 4| Broadband QLED and proposed white-light QLED. a**, Summary of correlation between $V_p$ of QDs in QLEDs and operating voltages of the devices at current density of 1 mA/cm² (V@1 mA/cm²). Top: the photographs of working QLEDs emitting different colors with a scale bar of 5 mm. **b**, Electroluminescence spectra of an orange-red βQLED produced using a blend of three types of QDs. Dashed lines show individual contributions from the three types of QDs comprising the emissive blend. **c**, CIE-*x* and *y* values of the βQLED in **b** as a function of luminance. **d**, Proposed color-mixed white-light QLED composed of three βQLEDs achieving a CRI of 90.4 referring to the black-body-radiation (BBR) at 3500 K. Inset: CIE-*x* and *y* values of three βQLEDs and white-light QLED.

# Extended data figures and tables

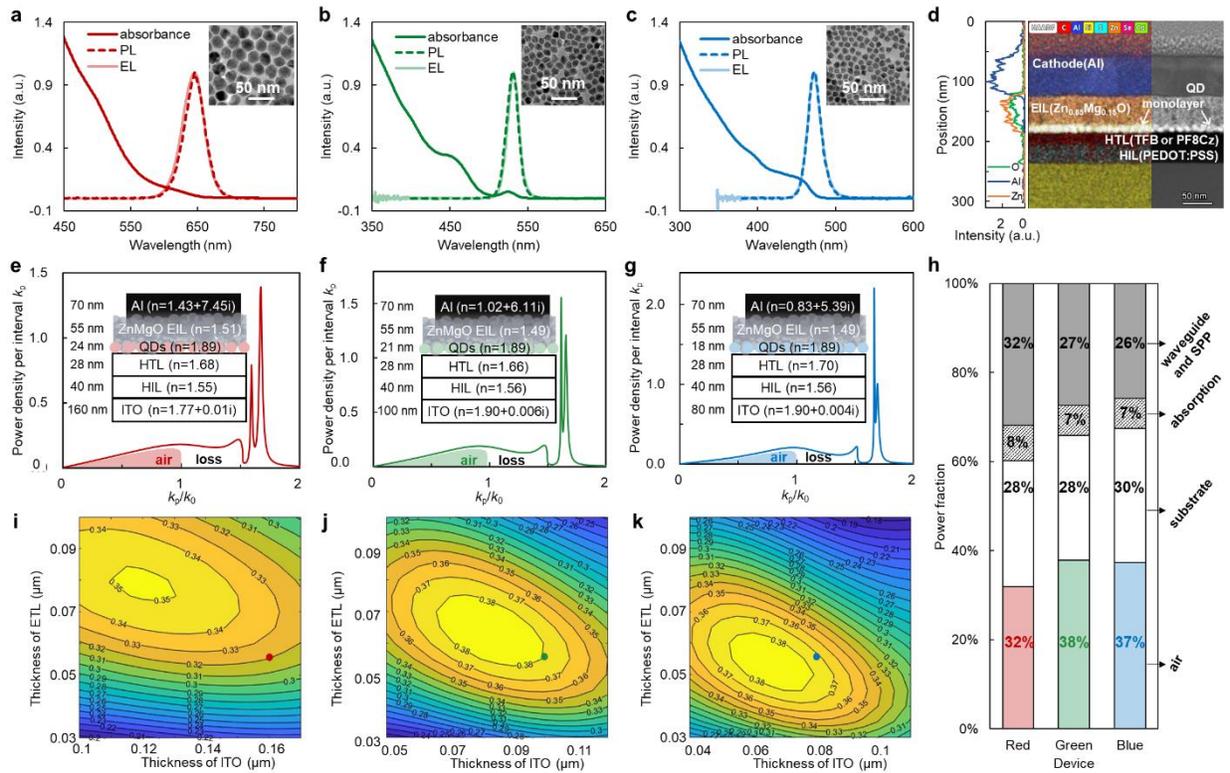

**Extended Data Fig. 1| Characterization of typical CdSe/Cd$_x$Zn$_y$Se$_z$S/ZnS core/shell/shell QDs and QLEDs, along with optical modeling of QLEDs.**

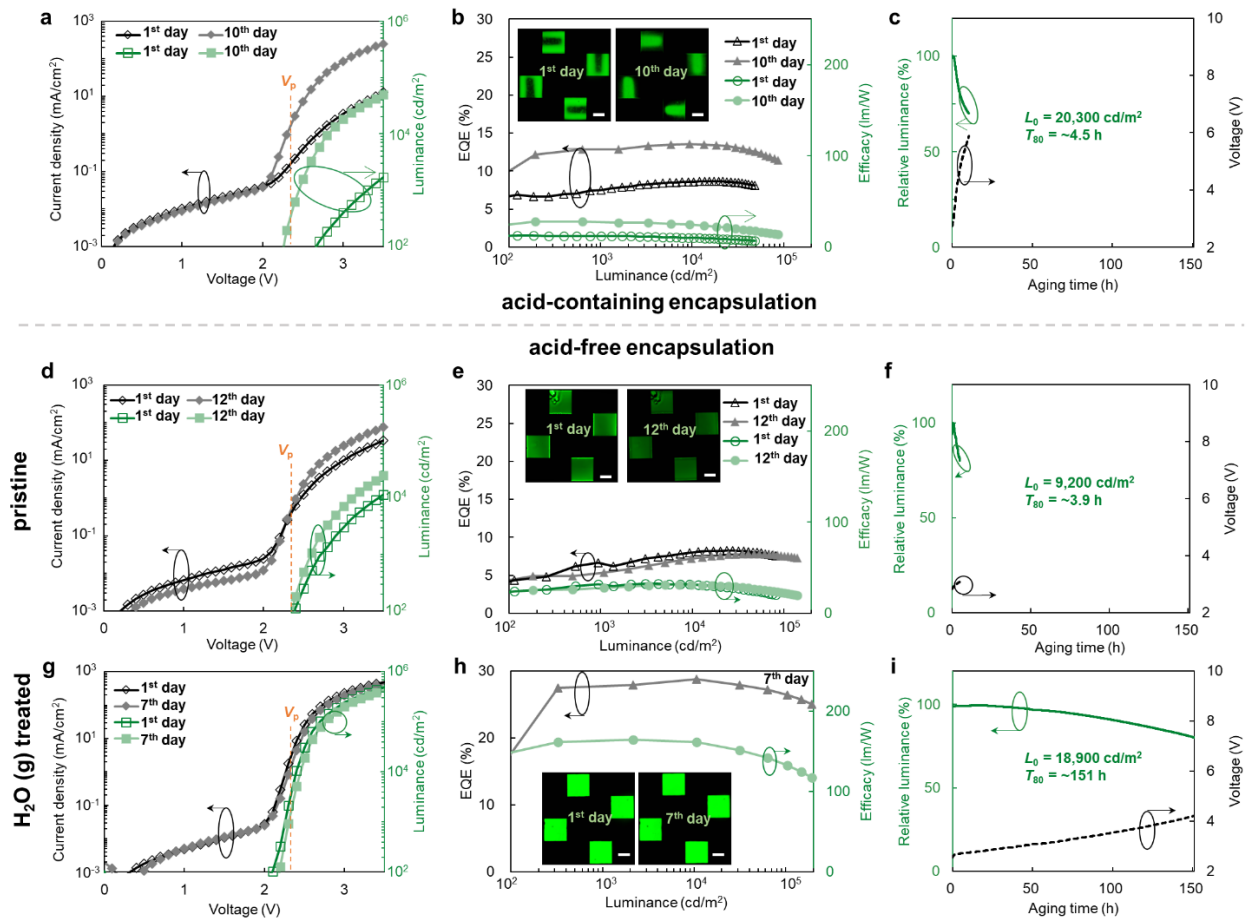

**Extended Data Fig. 2| Performance comparisons of green-emitting (532 nm) QLEDs encapsulated with the acid-containing resin (LOCTITE 3492), acid-free resin (LOCTITE 3335), and typical QLEDs with water-vapor treatment and encapsulated with the acid-free resin.**

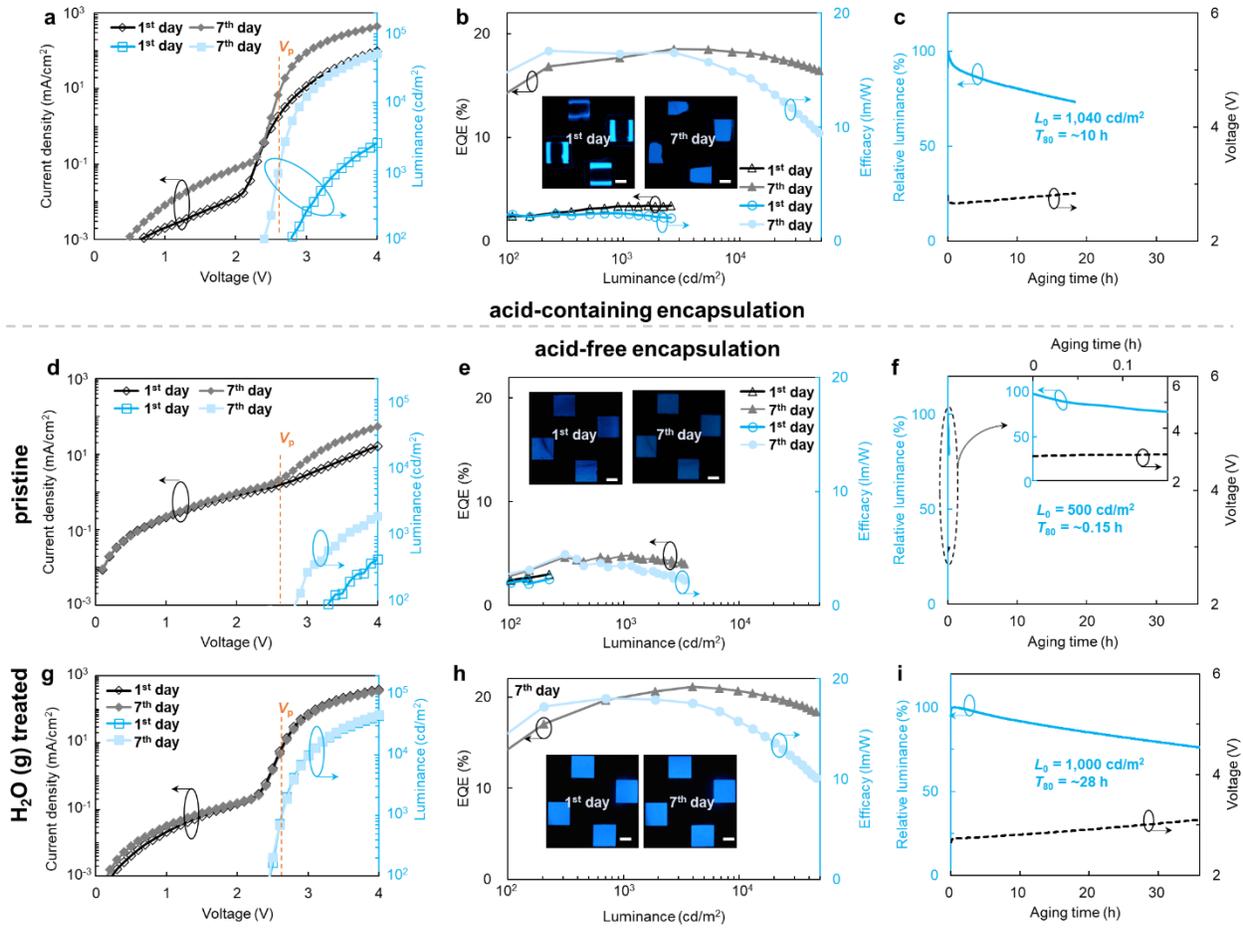

**Extended Data Fig. 3| Performance comparisons of blue-emitting (474 nm) QLEDs encapsulated with the acid-containing resin (LOCTITE 3492), acid-free resin (LOCTITE 3335), and typical QLEDs with the water-vapor treatment and encapsulated with the acid-free resin.**

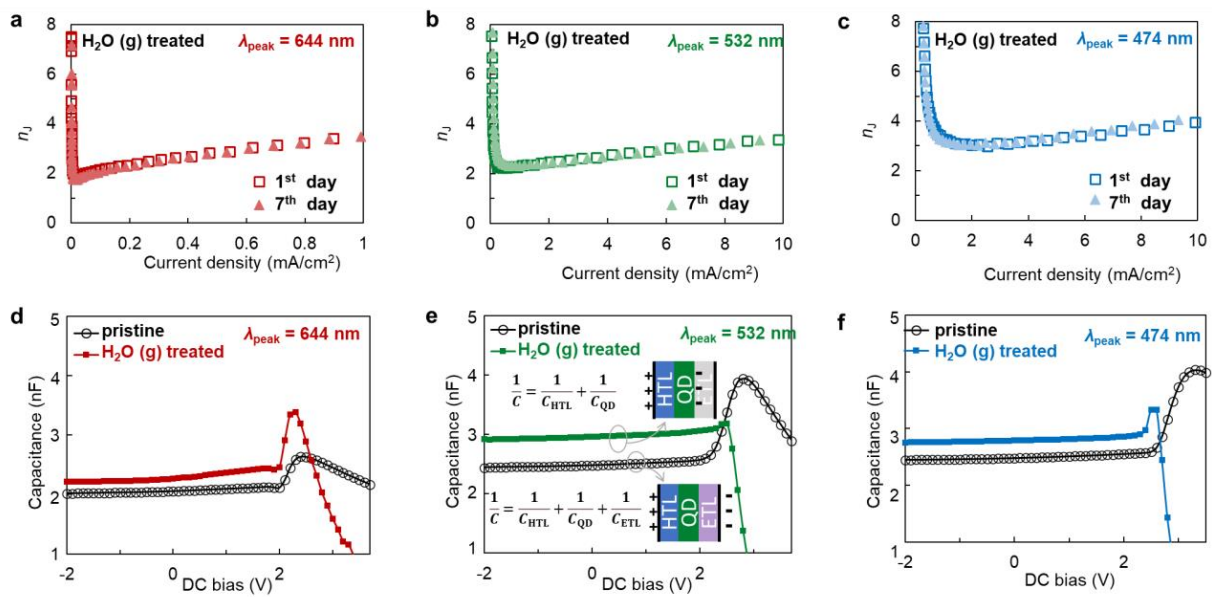

**Extended Data Fig. 4| Electronic properties of the QLEDs**. **a-c**, Temporal evolutions of current ideality factors ($n_J$) across various current densities. **d-f**, The capacitance-voltage ($C$–$V$) characteristics of QLED with ($H_2O$ (g) treated)/without (pristine) the water-vapor treatment.

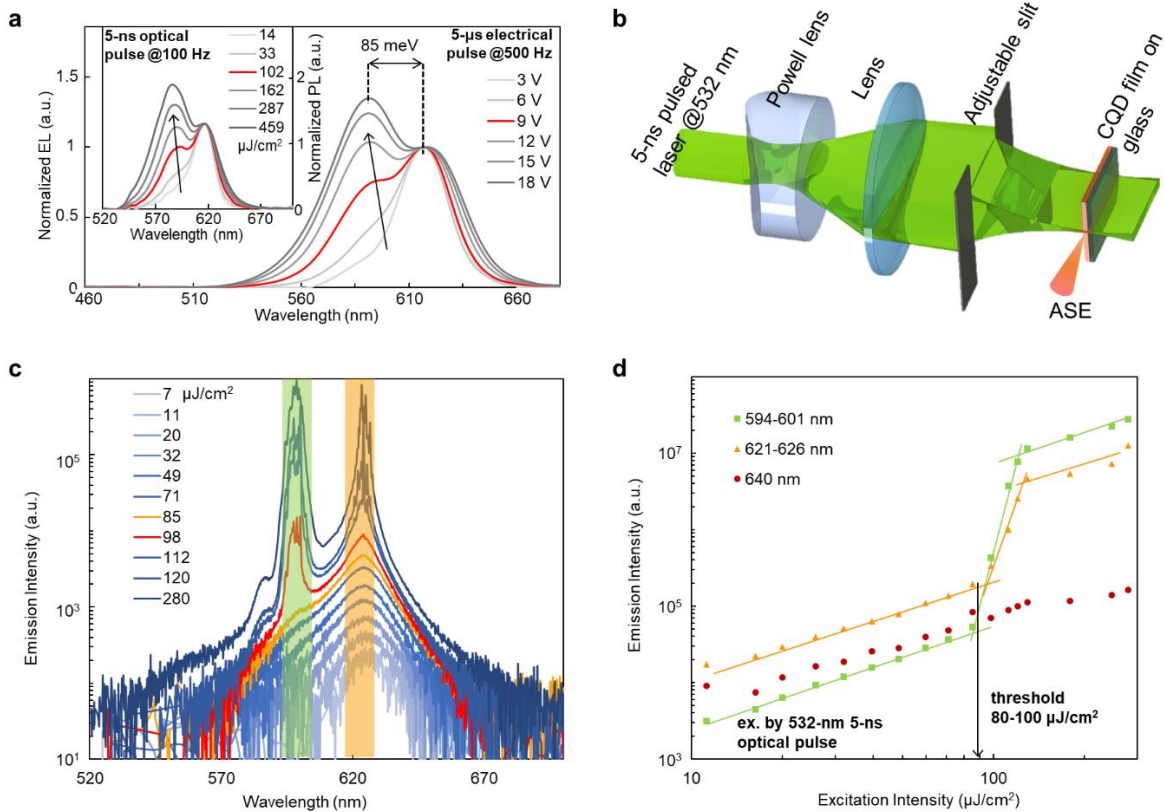

**Extended Data Fig. 5| Achievement of population inversion in red-emitting QLEDs confirmed by pulsed optical excitation.**

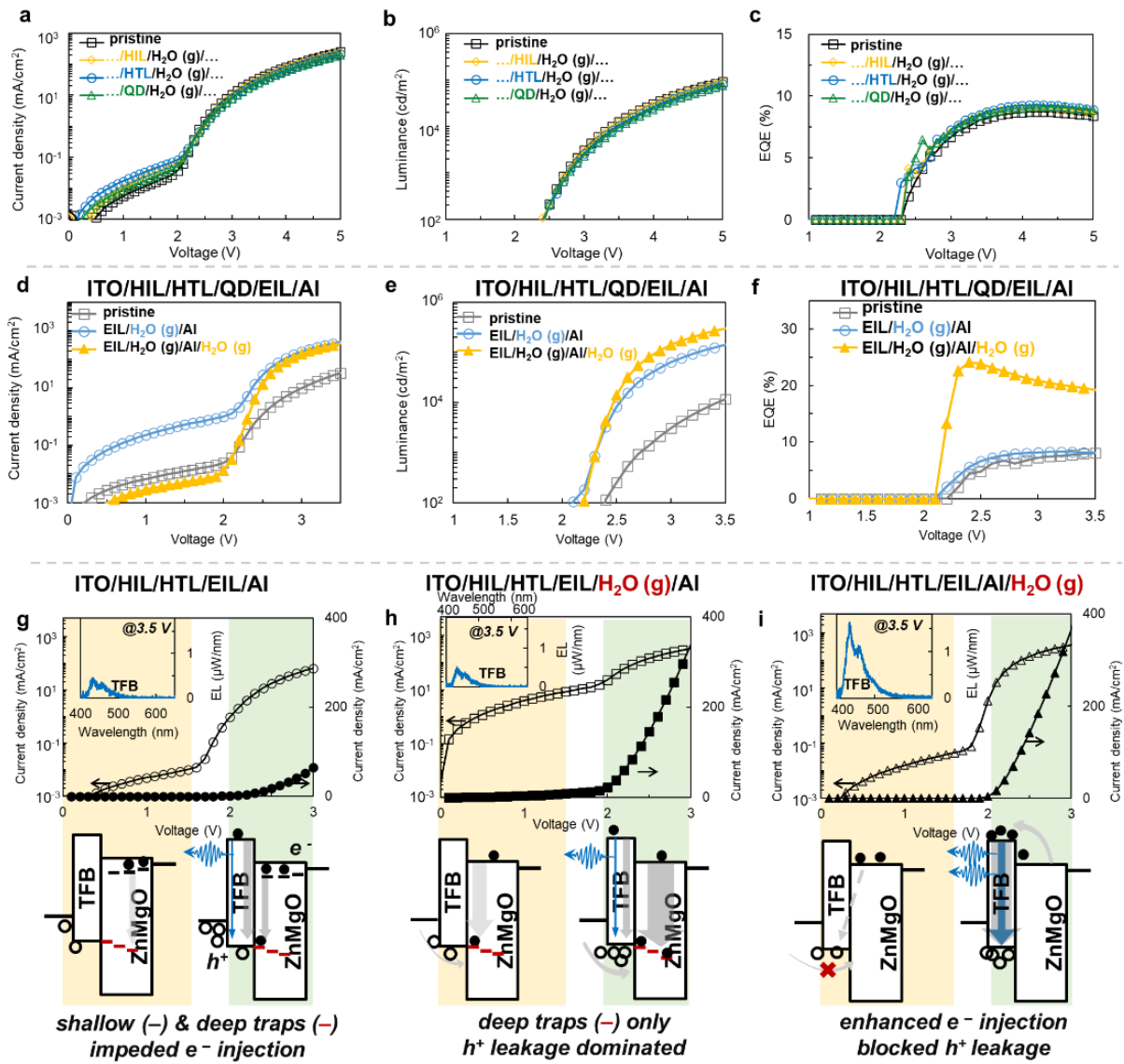

**Extended Data Fig. 6| A water treatment for various devices.**

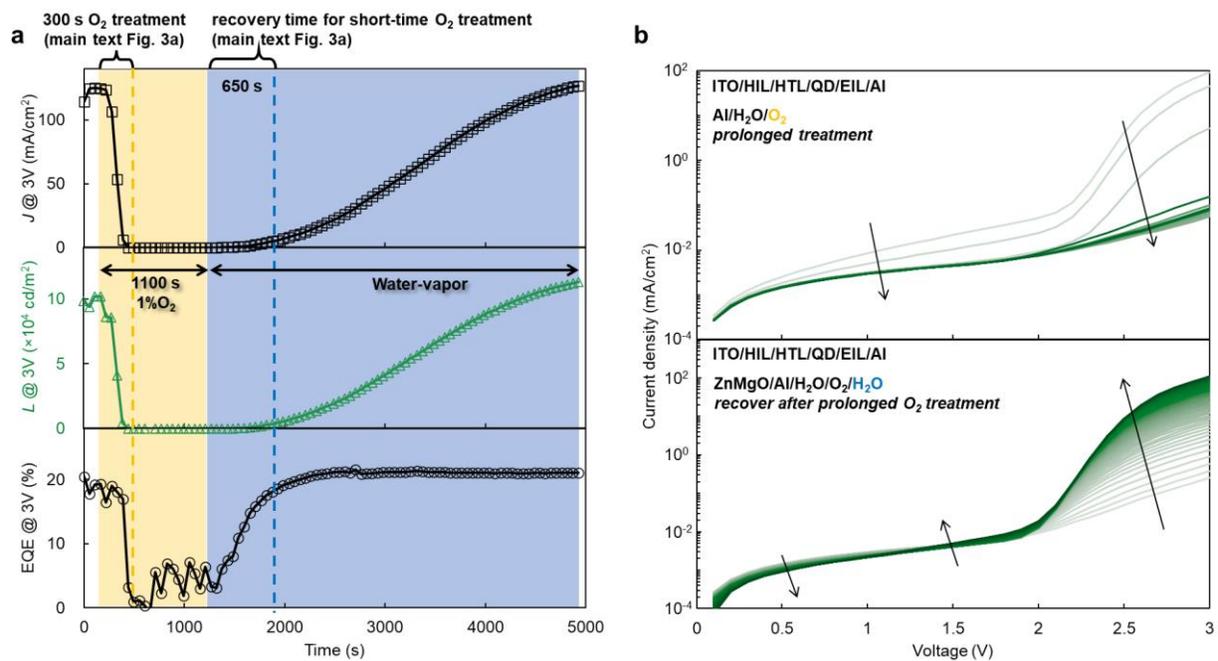

**Extended Data Fig. 7| An elongated oxygen treatment makes recovery much slower.**

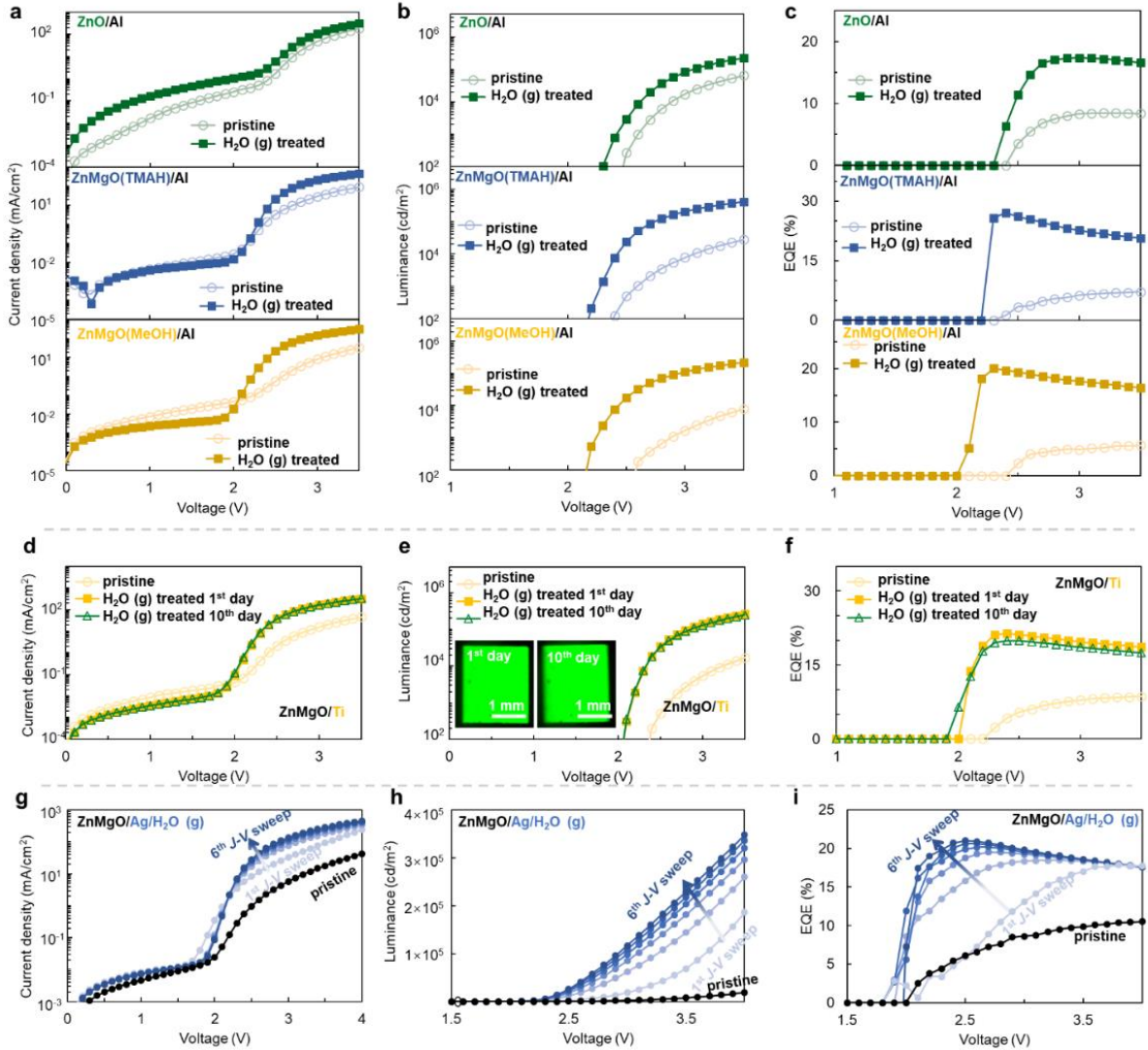

**Extended Data Fig. 8| Performance of the green-emitting (532 nm) QLEDs with different electron injection layers and top electrodes.**

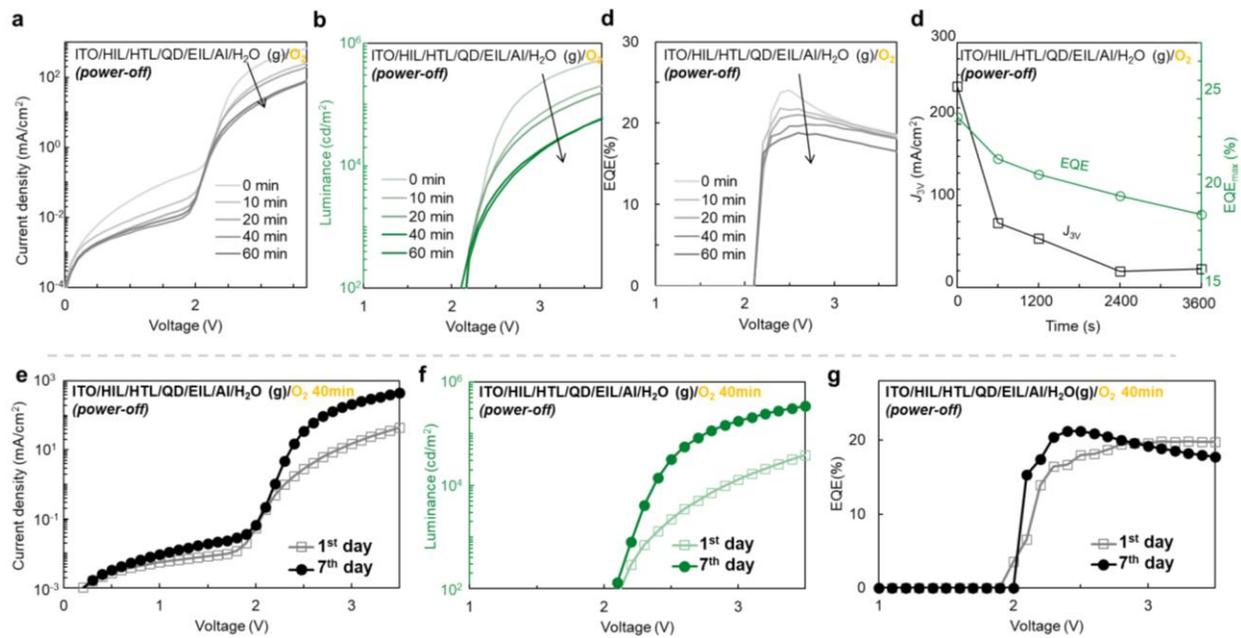

**Extended Data Fig. 9| Weak performance degradation and spontaneous recovery of green-emitting (532 nm) QLEDs after the power-off oxygen-treatment for different durations.**

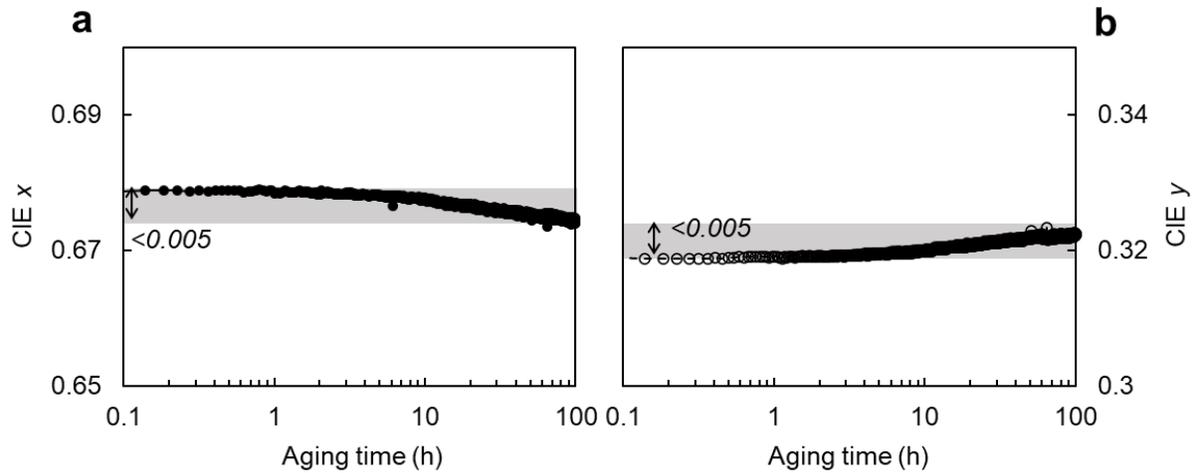

| Color | $\phi_L$(lm) | $\phi_L$ ratio | S (m²) | L (cd/m²) | $P_E$ ratio | $V_p$ (V) | $\eta_L$ (lm/W) | EPE (%) |
|---|---|---|---|---|---|---|---|---|
| Red-orange | 1160 | 29% | 0.24 | 1539 | 46.6% * (40.2%)** | 2.017 | 312 | 55 |
| Yellow-green | 2280 | 57% | 0.24 | 3025 | 28.1%* (29.0%)** | 2.268 | 596 | 46 |
| Green-blue | 560 | 14% | 0.24 | 743 | 25.3% * (30.8%)** | 2.570 | 159 | 32 |

$\phi_L$: luminous flux, L: luminance, S: emitting area, $\eta_L$: luminous efficacy, $P_E$: electrical power
*external power efficiency (EPE) = 100% is assumed
**certificated EPEs (last column) in the spectral range are taken into consideration

**Extended Data Fig. 10| a** and **b**, The color drift of $\beta$QLED during 100 hours continuous operation under constant bias of 1.90 V. **c**, Key parameters used in luminous efficacy estimation of the proposed white-light QLED.